\documentclass{INTERSPEECH2023}
\usepackage{todonotes}
\usepackage{booktabs}
\usepackage{multirow}
\usepackage{subcaption}
\usepackage{algorithm}
\usepackage{algpseudocode}
\newcommand{\norm}[1]{\left\lVert#1\right\rVert}

\interspeechcameraready

\title{There is more than one kind of robustness: Fooling Whisper with adversarial examples}
\name{Raphael Olivier$^1$, Bhiksha Raj$^1$}
\address{
  $^1$Carnegie Mellon University, USA}
\email{first@university.edu, second@companyA.com, third@companyB.ai}

\email{\{rolivier,bhiksha\}@cs.cmu.edu}
\begin{document}

\maketitle
%
\begin{abstract}
Whisper is a recent Automatic Speech Recognition (ASR) model displaying impressive robustness to both out-of-distribution inputs and random noise. In this work, we show that this robustness does not carry over to \textit{adversarial} noise. We show that we can degrade Whisper performance dramatically, or even transcribe a target sentence of our choice, by generating very small input perturbations with Signal Noise Ratio of 35-45dB. We also show that by fooling the Whisper language detector we can very easily degrade the performance of multilingual models. These vulnerabilities of a widely popular open-source model have practical security implications and emphasize the need for adversarially robust ASR.
\end{abstract}
\section{Introduction}
\label{sec:intro}
The improvements of Automatic Speech Recognition (ASR) models on academic benchmarks do not systematically lead to better performance in practical settings.
That requires another quality, commonly called \textit{robustness}. 
But different types of shifts between benchmarks and practical settings lead to as many definitions of robustness.
There is robustness to out-of-distribution data, i.e. the ability to generalize to different datasets than the ones used in training. 
There is robustness to random noise and in particular ambient noise. 
Finally, there is adversarial robustness, i.e. to perturbations generated by a third party with the specific intent to fool the model.

Recently the Whisper ASR model \cite{radford2022whisper}, a transformer sequence-to-sequence model trained on very large amounts of supervised data, was released in English-only and multilingual versions.
Its authors do not use data augmentation, relying solely on their large dataset to encourage robustness. 
Indeed, they achieve very impressive robustness against noise and out-of-distribution data. 
Whisper's remarkable performance has soon led to several applications like automatic captioning of online videos or speech-to-image generation models.

Its robustness to adversarial perturbations, however, has not yet been evaluated. 
It has been shown in past work that successful adversarial attacks on ASR models could lead to troubling security threats in multiple settings, in particular when the attacker has white-box access to the model \cite{abdullah2020sok}. 
Whisper is open-source: if it is vulnerable to adversarial attacks, then its deployment in real-world applications could turn these potential dangers into real liabilities.

In this paper, we show that Whisper is indeed vulnerable to white-box adversarial attacks. 
We use both targeted and untargeted attack algorithms, which we describe in section \ref{sec:method}.
We use them to modify inputs sampled from the LibriSpeech dataset \cite{librispeech} with near-imperceptible perturbations, achieving 35 to 40dB Signal-Noise Ratio (SNR) in most cases (section \ref{sec:exps}).
We show that these adversarial examples can fool Whisper models of all sizes for a large majority of inputs, either with untargeted attacks to mistranscribe predictions or with targeted ones to transcribe a specific target sentence of the attacker's choice (section \ref{sec:results}). 
Against untargeted attacks, Whisper is not any more robust than previously evaluated models. 
Against targeted attacks, multilingual models resist a little better, but can still be fooled by most inputs (\ref{sec:asrattackres}).
In addition, multilingual Whisper models can easily be fooled to detect the wrong language: this very simple attack is sufficient to entirely mistranscribe sentences in low-resource languages (\ref{sec:langattackres}). We propose a \textit{universal} version of this attack, with a single perturbation that can fool multiple speech utterances.
In section \ref{sec:analysis} we finally discuss the practical implications and limitations of these results. We also combine Whisper with a state-of-the-art ASR defense and recover some robustness at the cost of an important performance tradeoff (9\% WER on LibriSpeech test-clean).

To summarize, by studying Whisper this work makes two novel contributions: (1) we show that using massive amounts of diverse data does not robustness in the adversarial regime, contrary to the random noise regime; and (2) we propose a novel attack setting to specifically fool multilingual models on low-resource settings.
We release our code and all our adversarial examples alongside this paper\footnote{\url{https://github.com/RaphaelOlivier/whisper_attack}}.
We invite the speech modeling community to remain cautious when applying ASR models to security-critical settings and to combine them with adversarial defenses when doing so.

\section{Related work}
\label{sec:relwork}
Adversarial attacks were first proposed on image classification \cite{goodfellow13, goodfellow2014}, and were later extended to ASR models \cite{houdini,carlini18,qin19}.
Most works have focused on HMM or RNN-based models, but a few recent papers have compared the vulnerabilities of different neural architectures \cite{lu21c_interspeech, Olivier2022RI,sslspeechadv}, though none on Whisper.

The interactions between robustness to adversarial perturbations and to other types of noise have not been explored for ASR so far. 
They however have been studied for image classification. 
Some works have unified adversarial and random noise under a \textit{semi-random} regime and derived theoretical robustness bounds \cite{Fawzi16,rice21}. 
Moreover, a model robust to white noise can be indirectly useful in adversarially robust pipelines \cite{Lecuyer19,cohen19}.

We also study specific attacks on multilingual ASR models. Other works have shown the fragility of multilingual language models under perturbations of text inputs \cite{Rosenthal2021AreMB}. 
For ASR, some past work has compared the robustness of monolingual models trained in different 
languages \cite{Markert_2021}.

\section{Adversarial attacks}
Adversarial attacks consist of generating input perturbations designed to manipulate model outputs while evading human detection. 
We focus on norm-based white-box attacks: using model gradients for optimization, we generate an additive perturbation $\delta$ to the input $x$ such that $\norm{\delta}_p$ is small. 
We control the perturbation size with the Signal-Noise Ratio $\text{SNR}(\delta,x) = 20(\log\norm{x}_2-\log\norm{\delta}_2)$, expressed in decibels (dB).

Adversarial attacks differ in their objectives. 
We distinguish \textit{untargeted} attacks which prevent models from predicting the correct output; and \textit{targeted} attacks which fool models into predicting a specific target chosen by the attacker. 
On ASR models, this means transcribing a specific sentence. 
Untargeted attacks, which are satisfied by predicting random sentences or gibberish, are much easier.
We apply both types of attacks, using different algorithms due to the difference in attack difficulty.
\label{sec:method}
\subsection{PGD attack}
We use the Projected Gradient Descent (PGD) attack \cite{madry18} in $L_p$ loss, with $p=2$ and $p=\infty$. 
We optimize $\delta$ under the following constrained objective:
$$\max_{\norm{\delta}_p<\epsilon} \mathcal{L}(f(x+\delta),y)$$
where $f$ is the model (Whisper), $x$ the input, $y$ its correct transcription and $\mathcal{L}$ the loss function (cross-entropy). 
To keep $\delta$ in the $L_p$-ball of radius $\epsilon$, the optimization algorithm is projected gradient descent for $n$ steps: at each step, we run one gradient update, then if $\delta$  is too large we project it back onto the ball.
PGD can run targeted attacks, 
by minimizing the loss with the target $y_t$ instead.
However, targeted ASR attacks require additional optimization tricks. 

\subsection{Modified CW attack}
We use a version of the targeted Carlini\&Wagner (CW) attack \cite{carlini16,carlini18}: with target sentence $y_t$ the optimization objective is 
$$\min_{\norm{\delta}_\infty<\epsilon} \mathcal{L}(f(x+\delta),y_t) + c\norm{\delta}_2^2$$
i.e. we use both a $L_\infty$ bound (by clamping $\delta$ after optimization steps) and a $L_2$ regularization term with coefficient $c$.
We set a fairly large initial $\epsilon$, then gradually decrease it with the following schedule. 
At every optimization step, we check whether the model transcribes the target $y_t$. 
If it does, we multiply $\epsilon$ by a factor $\alpha<1$ and keep optimizing. 
We run this algorithm for at most $n$ optimization steps while decreasing $\epsilon$ at most $k$ times.

We found that the vanilla CW attack has trouble fooling Whisper. While investigating why that is the case, we observed that the first predicted token is particularly hard to push toward the target.
Therefore, we strengthen its coefficient in the aggregated loss over a sequence of length $L$, by setting: 
$$\mathcal{L}(f(x),y_t) = \frac{1}{L+\lambda}[(1+\lambda)\mathcal{L}(f(x)_1,{y_t}_i) + \sum_{i= 2}^L\mathcal{L}(f(x)_i,{y_t}_i)]$$
We find $\lambda=1$ to work well in practice. All CW results that we report thereafter use this modified attack. 

\subsection{Language confusion attack}
Multilingual Whisper models run a language detection module before transcribing sentences (see Section \ref{sec:whisper}).
Alongside attacking the decoder directly, we investigate the consequences of attacking that language detector, which is essentially a classifier trained with cross-entropy loss. 
We use the PGD targeted attack to push the prediction from the original language to another target language. We evaluate how this affects ASR performance.

Given that this is a simple attack objective, we also try applying a more restrictive threat model and run a \textit{universal} attack. 
We optimize a single $\delta$ to fool not just one but all inputs $x$. 
Specifically, we train a 30-second long parameter $\delta$ to fool the following objective: 

$$\max_{\norm{\delta}_p<\epsilon} \mathbb{E}_{x\in\mathcal{D}}\mathcal{L}(f(x+\delta),y)$$
To optimize it we combine utterances into a small "training set" $\mathcal{D}$, and train with PGD for several epochs. We then evaluate how that perturbation affects ASR performance on a test set disjoint from this training set.

\section{Experimental setting}
\label{sec:exps}

\begin{table*}[ht]
    \centering
    \begin{tabular}{llll|ll|ll|ll}
        Model & Params. & Clean & WN 0dB & \multicolumn{2}{c|}{$L_2$ PGD WER}  &  \multicolumn{2}{c|}{$L_\infty$ PGD WER} & \multicolumn{2}{c}{Carlini\&Wagner} \\
      & & WER & WER &  $\text{SNR}=35\text{dB}$ & \text{SNR}=40\text{dB}   & $\epsilon=5e^{-3}$ & $\epsilon=1.5e^{-3}$  & Acc. & SNR  \\
      \hline
      SpeechBrain  & 165M &  2.4\% & 89.1\% &  75.9\%   & 63.2\% &  93.0\% & 81.6\% &  100\% & 37dB \\
      \hline
        tiny.en  & 39M &  3.4\% &89.6\% & 99.0\%  & 87.8\& &  115\% & 100\% &  94.1\% & 41.3dB \\
        base.en  & 74M & 3.0\% & 64.7\% & 92.8\%  &  81.7\% & 104\% & 98.3\%  & 100\% & 38.7dB \\
        small.en   & 244M &  1.9\% &46.1\% &  67.2\% & 53.9\% & 78.7\% & 66.0\%   &  82.4\%  & 40.0dB\\
        medium.en   & 769M &  1.7\% & 40.3\% &  52.9\%  & 39.3\% & 65.2\% & 51.4\% & 76.5\% & 40.5dB \\
        \hline
        tiny  & 39M &  5.8\% &93.2\% &   103\%  & 96.1\% & 107\% & 104\% &    82.3\% & 36.6dB \\
        base  &  74M & 3.4\%& 77.9\%  & 90.3\% & 80.4\% & 102\%  & 96.7\% & 76.5\%   & 35.1dB \\
        small  & 244M  & 2.0\% & 52.0\% &   75.4\% &  61.5\% &  89.8\% & 77.4\% &  52.9\% & 32.0dB \\
        medium  &  769M  & 1.5\% & 36.5\% &  49.5\%  & 38.9\%   &  63.3\% & 49.3\% & 64.7\%  & 29.5dB \\
        large   & 1550M & 1.6\% & 32.6\% & 45.3\% & 34.1\% &   49.2\% & 38.8\% & 82.4\% & 26.6dB \\
      \hline
    \end{tabular}
    \caption{Results on LibriSpeech test-clean of the white-box attacks. For untargeted PGD attacks, we report the achieved WER with a 35dB and 40dB respective SNR for $L_2$ attacks, and with $\epsilon=0.005$ and $\epsilon=0.0015$ for $L_\infty$ attacks. We also report the Word-Error Rate on the same data under random white noise (WN). For the targeted CW attack, we report the proportion of successful adversarial examples (those that Whisper transcribes as the target) and their average SNR. We evaluate the Speechbrain transformer model (first line) and all Whisper models.
    \label{tab : asrattack}}
\end{table*}
We run all our experiments using only one Tesla A100 GPU.
We run attacks using SpeechBrain \cite{speechbrain} and robust\_speech \cite{Olivier2022RI}, within which we integrate the Whisper inference package provided by OpenAI, adding loss computation functions to it.

\subsection{Whisper models}\label{sec:whisper}
Whisper exists in 5 model sizes. We run our attacks on all models from tiny (39M parameters) to large (1550M). 
All models are Transformer sequence-to-sequence models, with an encoder turning speech inputs into contextual representations, and a decoder mapping them to language tokens. 
There are four English-only models trained on 438kh of supervised English training data, and five multilingual models trained with an additional 243kh of multilingual data.
Initial tokens in the decoder can specify the task and the language. 
A language detection module generates the latter if it is not specified and if the model is multilingual. We do not change any of the default inference hyperparameters: for example, the beam size is 5 for all models.

\subsection{Datasets}
To attack the ASR decoder we perturb inputs sampled from the LibriSpeech test-clean dataset: 75 for the untargeted attack and 17 for the (much longer) targeted attack. 
As target transcription for the CW attack, we follow the example of \cite{carlini18} and use the fixed sentence "OK Google, browse to evil.com". 
This sentence is arbitrary and could be replaced with any other.

To attack language detection, we use the multilingual CommonVoice dataset \cite{commonvoice}. We sample 100 sentences from the test set of each of the following seven languages: Armenian, Lithuanian, Czech, Danish, Indonesian, Italian, and English. 
Whisper was trained on varying amounts of these languages, making them representative of the distribution of its training data. 
We use three target languages: English, Tagalog, and Serbian, also present in very different amounts in the Whisper training set. For the universal language confusion attack, the attack training set consists of 70 sentences (10 per source language), disjoint from the seven test sets, and the target language is Serbian.

\subsection{Attack hyperparameters}
For the PGD attack in $L_2$ norm, we set SNR objectives of 30 and 40dB, which correspond to very small noise. 
We then compute for each utterance the corresponding $\epsilon$ bound. 
We use an attack learning rate of $0.1*\epsilon$, and $n=200$ attack iteration steps. This attack on Whisper medium runs in 2 minutes on one A100 for a typical utterance. For the $L_\infty$ attack we fix  $\epsilon=0.005$ or $\epsilon=0.0015$ (on average SNRs of 38dB and 49dB respectively).
For the CW attack, we use $n=2000$ iteration steps ($\sim$25min for Whisper medium), the Adam optimizer with learning rate $0.01$, and regularization term $c=0.25$ for the tiny and base models, $c=1$ for larger models.
Our initial radius is $\epsilon=0.1$: this corresponds on average to a 15dB SNR. 
We decrease $\epsilon$ up to $k=8$ times by a factor $\alpha=0.7$ when the model predicts the target, for a final SNR of up to 45dB.

For the white-box language confusion attack, we set an SNR of 45dB and $n=30$ iteration steps. The universal attack uses a SNR of 40dB, and fits $\delta$ for 2000 epochs over the 70 training sentences using one iteration step per input and a learning rate of $0.001*\epsilon$


\subsection{Metrics}
Untargeted and language attacks aim at degrading ASR performance. 
We evaluate them with the Word-Error Rate (WER) on correct transcriptions. 
For targeted attacks, the goal is to predict perfectly the attack target with little noise. 
Therefore, we evaluate them with sentence-level accuracy, i.e. we only consider an attack successful if Whisper transcribes the target exactly. 
Moreover, we only consider attacks successful if the achieved SNR is higher than 30dB.
\section{Results}
\label{sec:results}

\begin{figure*}[ht]
     \centering
         \includegraphics[width=0.75\textwidth]{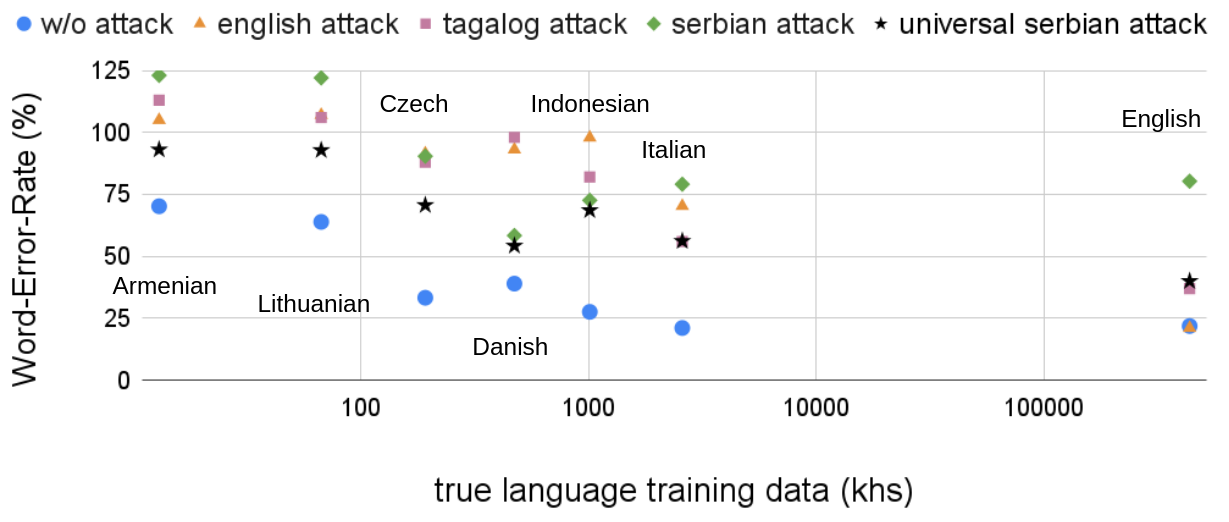}
         \caption{WER of the multilingual Whisper medium on subsets of the CommonVoice dataset in 7 languages. We fool the model with white-box attacks into wrongfully detecting either English, Tagalog, or Serbian. We also fool it with a universal attack to predict Serbian. The x axis corresponds to the amount of training data Whisper was trained on for the \textit{true} language of the inputs.}
        \label{fig:plot}
\end{figure*}
\subsection{Attack on ASR decoding} \label{sec:asrattackres}

In Table \ref{tab : asrattack} we report the results of the untargeted PGD and targeted CW attack. 
We observe that even with small perturbations, the PGD attack is largely successful in degrading the performance of all Whisper models, by 35 to 89\% in absolute WER for 40dB SNR and 48 to 99\% for 35dB SNR. 
Such degradation can prevent Whisper from being useful in most practical applications (e.g. video captioning with a 45\% WER is unsatisfying). 
$L_\infty$ PGD attacks are equally or more successful, with very small radii (in practice similar to SNRs of 38-49dB). 

For comparison, we run the same attacks on the SpeechBrain Transformer model \cite{speechbrain} trained on LibriSpeech,  using the same beam size of 5 during inference. We observe that attack results are very close to those of the Whisper small model, of similar size and clean performance. 
We can conclude from this that the Whisper training method does not lead to any improvement in adversarial robustness on untargeted attacks, compared to models of similar architecture trained in-distribution on fewer data. 
In contrast, Whisper models perform considerably better than this baseline under white noise, which we report in the same Table.
This illustrates the difference between robustness to \textit{average} (random) and \textit{worst-case} (adversarial) perturbations.


As for targeted attacks, in 50\% to 90\% of all sentences the CW attack succeeds in making the model predict the target with very little noise. 
This does show an increase in robustness compared to the baseline, but Whisper remains largely vulnerable to targeted attacks.
Models of all sizes can be fooled, but multilingual models are harder to attack. 
Most models are fooled with an SNR above 35dB; for the bigger multilingual models the SNR drops but remains above 25dB. 
This could indicate that multilingual training helps make models robust to targeted attacks, though not robust enough to evade most attacks.

\subsection{Language confusion attack on multilingual models}\label{sec:langattackres}
In Figure \ref{fig:plot} we plot the WER of the Whisper medium model under the language confusion attack, as a function of the amount of Whisper training data in the true language. 
We report our results for three possible attack targets: English, Tagalog, and Serbian.
We observe that in a large majority of cases, this very simple attack is sufficient to degrade significantly the WER of the ASR model.
When the true language is poorly represented in the Whisper training data, as for Armenian and Lithuanian, the WER jumps over 100\%. For higher-resource languages (Czech, Danish, Indonesian) the attack can degrade the WER to 75\% or higher.
For the best-represented languages, like English or Italian, the attack is a bit less effective but still degraded the performance by 16\% to 60\% absolute WER points. 
Intuitively, when the source and target languages are identical the attack has no effect, as observed with English.

We achieve this performance degradation with very low noise (45dB SNR) and an extremely simple attack with 30 iterations of PGD. 
This shows how brittle a multilingual model with language detection can be, especially over low-resource languages. 
While looking at the actual outputs on these low-resource languages under attack, we observed that the prediction is a mostly nonsensical mix of the true and target language.

Even the universal attack is largely successful in confusing Whisper, despite using a single perturbation for all inputs and all source languages. It achieves a Word-Error Rate degradation of 20 to 40\% for all languages. This perturbation can be used off-the-shelf on any input, without additional computation. With its SNR of almost $40dB$ it remains almost imperceptible.

The influence of the choice of target language on attack success is hard to derive from our results. The Serbian attack target has a stronger effect on most source languages, but Danish and Indonesian are exceptions. Studying whether linguistic proximity between source and target languages would be an interesting follow-up to this work.

\section{Implications and Mitigation}
\label{sec:analysis}
\subsection{Security threats on ASR and limitations}
The threat models we evaluate in this work cannot be applied to 100\% of ASR applications. 
Apart from the universal language confusion attack, whose WER degradation does not match white-box attacks, our adversarial algorithms need time to generate noise tailored for each input. Moreover, they modify inputs in the digital space, not under real-world acoustic conditions.
As a result, applying these attacks over-the-air as people speak rather than in the digital waveform space is not doable at this time.
However, other works have extended adversarial attacks to be generated over the air \cite{qin19,imperio19} and in real-time \cite{lu21c_interspeech,Xie2021EnablingFA}. Applications of those works to Whisper may be possible and would extend our results to many more threat models.

In addition, our simple threat models are sufficient to fool Whisper in several practical situations. For instance, if ASR is used to filter speech inputs, e.g. to remove hateful content from an online platform, new uploads can evade said detection with an untargeted attack. Other possible applications include \textit{censorship triggering} (using targeted attacks to fool Whisper into perceiving hateful content where there is not) or even \textit{data poisoning} if Whisper is used to generate new text corpus from audio.
With the recent improvements in ASR, such use cases are plausible enough to require caution from the speech community. 

\subsection{Defending Whisper with randomized smoothing}

\begin{table}[ht]
    \centering
    \begin{tabular}{llll}
     Model &   no attack & PGD 35dB & PGD 40dB  \\
    \hline
     undefended &  3.4\%  &  90.3\%  & 80.4\% \\
     $\sigma=0.02$ &  9.6\% &   35.6\% &  21.4\% \\
     $\sigma=0.03$ &   18.9\% &   37.6\% &  27.8\% \\
    \end{tabular}
    \caption{\label{tab:smooth} WER of the defended Whisper base model under PGD attack. 
    We use randomized smoothing with deviations $0.02$ and $0.03$. 
    We compare results to the undefended model.}
\end{table}

Fortunately, defenses against adversarial attacks exist. 
For example, randomized smoothing \cite{cohen19} is a simple defense that adds Gaussian noise over inputs before passing them to the model. 
It has been extended to ASR in the past \cite{zelasko,olivier-raj-2021-sequential}.

As the vanilla Whisper is fairly robust to Gaussian noise, it would likely be a good fit for this defense. We briefly verify it by using smoothing on the base model with Gaussian noise, using several standard deviations $\sigma$, and attacking it with PGD at SNR 35dB. The results, reported in Table \ref{tab:smooth}, show that smoothing mitigates the untargeted attack threat partially and recovers most of the degraded performance under attack. 
However, the performance tradeoff of this attack on unmodified inputs is significant, with a WER increase of 6 to 15 absolute points.

Using the larger models and further enhancements of the defense may improve those results.
Still, randomized smoothing is not a silver bullet.
Its robustness guarantees apply to norm-bound and specifically $L_2$-bound perturbations. 
This covers the $L_2$ PGD attack but not all adversarial attacks in the literature. 
Further research is required to achieve robust ASR.

\section{Conclusion}
\label{sec:ccl}
Despite its robustness to natural or random perturbations and distributional shifts, Whisper is highly vulnerable to adversarial examples. In particular, ASR performance in "low-resource" languages (relative to English) is very simple to degrade. Both targeted and untargeted white-box ASR attacks are effective as well. These vulnerabilities are the source of practical, concerning liabilities and emphasize the need for adversarially robust ASR models.

They also suggest interesting directions of progress for speech modeling. 
Whisper's out-of-distribution generalization, emerging from large amounts of training data, may not bring adversarial robustness;
but inversely, adversarially robust models do often generalize better to new domains than non-robust ones. 
Therefore extending adversarial training to ASR may yield a path toward even greater or less data-consuming generalization.
\vfill\pagebreak

\label{sec:refs}

\bibliographystyle{main}
\bibliography{main}

\end{document}